\documentclass[conference]{IEEEtran}
\IEEEoverridecommandlockouts
\usepackage{cite}
\usepackage{amsmath,amssymb,amsfonts}
\usepackage{algcompatible}
\usepackage{algorithm}
\usepackage{graphicx}
\usepackage{textcomp}
\usepackage{mathtools}   
\usepackage{lipsum} 
\usepackage{subfig}
\usepackage{amsfonts}
\usepackage{amssymb}
\usepackage{color}
\usepackage{caption}
\usepackage{mathtools}
\usepackage{relsize}
\usepackage{amsthm}
\usepackage{amssymb}
\usepackage{amsmath}
\usepackage{algpseudocode}
\usepackage{caption}
\usepackage{subcaption}
\usepackage{setspace}
\usepackage[colorlinks=true, allcolors=blue]{hyperref}
\def\BibTeX{{\rm B\kern-.05em{\sc i\kern-.025em b}\kern-.08em
    T\kern-.1667em\lower.7ex\hbox{E}\kern-.125emX}}
\begin{document}
\title{NL-CPS: Reinforcement Learning-Based Kubernetes Control Plane Placement in Multi-Region Clusters}

\author{\IEEEauthorblockN{Sajid Alam}
\IEEEauthorblockA{\textit{School of Computing, Engineering,}\\
\textit{and the Built Environment} \\
\textit{Edinburgh Napier University}\\
Edinburgh, UK \\
s.alam2@napier.ac.uk}
\and
\IEEEauthorblockN{Amjad Ullah}
\IEEEauthorblockA{\textit{School of Computing, Engineering,}\\
\textit{and the Built Environment} \\
\textit{Edinburgh Napier University}\\
Edinburgh, UK \\
a.ullah@napier.ac.uk}
\and
\IEEEauthorblockN{Ze Wang}
\IEEEauthorblockA{\textit{School of Computing, Engineering,}\\
\textit{and the Built Environment} \\
\textit{Edinburgh Napier University}\\
Edinburgh, UK \\
z.wang3@napier.ac.uk}
}

\maketitle
\begin{abstract}
The placement of Kubernetes control-plane nodes is critical to ensuring cluster reliability, scalability, and performance, and therefore represents a significant deployment challenge in heterogeneous, multi-region environments. Existing initialisation procedures typically select control-plane hosts arbitrarily, without considering node resource capacity or network topology, often leading to suboptimal cluster performance and reduced resilience. Given Kubernetes’s status as the de facto standard for container orchestration, there is a need to rigorously evaluate how control-plane node placement influences the overall performance of the cluster operating across multiple regions. This paper advances this goal by introducing an intelligent methodology for selecting control-plane node placement across dynamically selected Cloud–Edge resources spanning multiple regions, as part of an automated orchestration system. More specifically, we propose a reinforcement learning framework based on neural contextual bandits that observes operational performance and learns optimal control-plane placement policies from infrastructure characteristics. Experimental evaluation across several geographically distributed regions and multiple cluster configurations demonstrates substantial performance improvements over several baseline approaches.

\end{abstract}
\section{Introduction} \label{sec:intro}
Modern distributed cloud-native applications increasingly demand globally deployed services that are both low-latency and highly available~\cite{ref1}. Automating the deployment and management of such applications has led to the widespread adoption of orchestration solutions. Extensive research has explored high-level orchestration frameworks in both academia and industry nevertheless, Kubernetes has become the de facto standard for low-level orchestration of containerised workloads in distributed environments~\cite{ullah2023orchestration,bohm2022cloud,Tomarchio2020}.

Kubernetes employs a centralised architecture, where the core component, called control-plane---comprising the API server, etcd distributed datastore, scheduler, and controller manager---manages cluster-wide coordination and decision-making, exerting a profound influence on overall performance and operational efficiency~\cite{ref2}.
Lightweight Kubernetes distributions (e.g. K3S) consolidate all control-plane components into a single k3S-server, reducing memory consumption by approximately 50\% compared to standard Kubernetes, making it an ideal choice for edge computing, where heterogeneous resources and wide-area network conditions introduce non-trivial placement trade-offs. As a result, the initial placement of the control-plane within the resource stack is a critical architectural decision. 

Currently, placement is often determined manually or via simplistic heuristics without accounting for infrastructure characteristics. Consequently, the control-plane is often deployed on nodes that are resource-constrained or have unfavourable network positioning relative to the majority of worker nodes, leading to degraded API server responsiveness, reduced pod deployment throughput, and increased resource contention. Furthermore, as cluster size and geographic dispersion increase, manual placement decisions based on operator intuition scale poorly, underscoring the need for automated and intelligent control-plane placement strategies. This need is further strengthened in dynamic environments, where the resource stack is selected at runtime by an orchestration framework (e.g., as in Swarmchestrate~\cite{UllahMAKKDMWK25,renau2025distributed}) to meet QoS objectives, where manual control-plane placement is no longer feasible.

We frame the control-plane placement problem as a \textit{contextual bandit} learning task, in which the system observes infrastructure characteristics (context), selects a host node (action), and receives operational performance feedback (reward) within a single decision cycle. This represents a single-step decision problem, i.e. once the control-plane is deployed and the cluster is initialised, the episode terminates without further state transitions or temporal dependencies. This distinguishes our problem from sequential reinforcement learning scenarios, such as dynamic workload scheduling~\cite{xing2025self}, adaptive container migration~\cite{zhu2024adaptive}, or multi-stage resource allocation~\cite{liu2017hierarchical}, where agents must navigate sequences of states influenced by prior actions. More specifically, our key contributions are as follows.
\begin{itemize}
   \item We formalise K3S control-plane placement as a contextual bandit problem, thereby establishing theoretical foundations distinct from sequential workload scheduling.
   \item We propose Neural LinUCB for control-plane node Selection (NL-CPS), a framework that combines neural network function approximation with UCB-driven exploration to achieve balanced exploitation-exploration.
   \item We develop a synthetic training environment calibrated from real K3S measurements across a wide-range of cluster configurations, enabling rapid policy learning without repeated physical deployments.
   \item We conduct extensive evaluation on real 12-node and 18-node multi-region deployments spanning 10 geographic regions, demonstrating that policies trained on synthetic data generalise effectively to unseen cluster topologies.
\end{itemize}
The remainder of this paper is structured as follows. Section~\ref{sec:related} provides background on the Swarmchestrate orchestration framework and reviews related work. Section~\ref{sec:problem} formalises the problem and presents 
the NL-CPS framework. Section~\ref{sec:evaluation} describes the 
experimental setup and the evaluation results. Finally, Section~\ref{sec:Conc} concludes the paper and outlines future work.


\section{Background, Motivation, and Related Work} \label{sec:related}
This section first provides an overview of the distributed orchestration framework---Swarmchestrate~\cite{kiss2024swarmchestrate}---within which our proposed mechanism operates, followed by a review of related work.
\subsection{Swarmchestrate overview} \label{sec:swarmchestrate}
Swarmchestrate, an ongoing Horizon Europe project, is a distributed orchestration framework for the cloud-edge continuum. Its vision is to enable self-organising capabilities that support dynamic, context-aware deployment and management of complex applications across heterogeneous infrastructure, without a central orchestrator. More specifically, it targets three key challenges: first, enabling seamless, concurrent access to a highly heterogeneous resource landscape while coordinating end-to-end services across multiple cloud, fog, and edge providers. Second, optimising resource selection to balance key objectives such as performance, cost, and energy efficiency and third, ensuring trust, reliability, and security in deployments spanning diverse administrative domains, geographic locations, and network environments. This paper contributes to the first challenge.

The Swarmchestrate interface is a P2P network of \textit{Resource Agents (RA)}, each representing a distinct resource provider with access to and local knowledge of its own set of resources. RAs can dynamically join or leave the network and are primarily responsible for exposing their resources and collaborating with other RAs to discover and allocate suitable resources from their capacities to incoming applications. Swarmchestrate supports microservices-based applications, each comprising $m$ components with specific resource requirements. Upon submission, the application request is broadcast to all RAs, who then initiate a collaborative process to identify and reach consensus on the most suitable resources (refer to~\cite{renau2025distributed}). Once the set of resources for a given application has been identified, these resources are dynamically instantiated to form a Kubernetes cluster responsible for application deployment and runtime management. This paper focuses on the specific challenge of dynamically selecting an appropriate resource to host the Kubernetes control-plane from among the identified resources. For a detailed description of the Swarmchestrate architecture and the integration of this work, we refer the reader to~\cite{UllahMAKKDMWK25}, while its broader vision is discussed in~\cite{kiss2024swarmchestrate}.

\subsection{Related Work} \label{sec:related_work}
This section reviews related work across the following four key areas. Table~\ref{tab:related_work} further summarises the positioning of our work relative to existing research.

\subsubsection{Lightweight Kubernetes Distributions}
A comprehensive performance comparison of several lightweight distributions (including MicroK8s, K3s, K0s, and MicroShift) is carried out in ~\cite{koziolek2023}, evaluating control-plane throughput, data plane latency, and resource consumption. A similar evaluation of Kubernetes distributions is also carried out in ~\cite{kjorveziroski2022}, however, for serverless workloads at the edge. Böhm and Wirtz~\cite{bohm2021} profiled the lifecycle resource consumption of MicroK8s and K3S. Čilić \textit{et al.}~\cite{cilic2023} evaluated container orchestration tools including K3S and KubeEdge in edge environments, assessing deployment complexity, 
memory footprint, and network-aware performance. While these studies provide valuable insights into the different Kubernetes lightweight distribution characteristics, none address the problem of optimal control-plane node selection and its impact on overall cluster performance in heterogeneous, multi-region deployments. 

\subsubsection{Kubernetes optimisation techniques}
Recent research has increasingly focused on optimising Kubernetes performance. For example, Mondal \textit{et al.}~\cite{mondal2024kubernetes} addressed control-plane optimisation, including ETCD performance, backup strategies, and scheduling algorithms. Load-aware orchestration strategies~\cite{loadaware2025} have been proposed to enhance pod placement decisions based on runtime cluster state. In the 
multi-cluster domain, frameworks such as Karmada~\cite{karmada} have emerged for cross-cluster orchestration. While these systems address workload distribution across clusters, they do not optimise the initial control-plane placement decision within individual clusters.
\subsubsection{Reinforcement Learning for Kubernetes Scheduling}
\label{sec:rl_k8s}
Several research works have explored reinforcement learning approaches for Kubernetes scheduling decisions. For example, Huang~\textit{et al.}~\cite{huang2020rlsk} proposed RLSK, a deep reinforcement learning-based job scheduler for federated Kubernetes clusters that learns scheduling strategies without prior knowledge of the multi-cluster environment. Han~\textit{et al.}~\cite{han2021kais} introduced KaiS, a learning-based 
scheduling framework for Kubernetes-based edge-cloud systems that combines graph neural networks with multi-agent actor-critic algorithms to improve request dispatch and service orchestration. 
Jian~\textit{et al.}~\cite{jian2024drs} presented DRS, a deep reinforcement learning based Kubernetes scheduler that frames pod scheduling as a Markov decision process. 
Wang~\textit{et al.}~\cite{wang2023ppolrt} proposed PPO-LRT for edge K3S clusters, using proximal policy optimisation to balance cluster load during task deployment. A comprehensive survey in~\cite{carrion2022survey} 
categorises Kubernetes scheduling algorithms and identifies 
AI-based approaches are a growing research direction. However, 
all existing RL-based approaches target \textit{dynamic pod 
scheduling}---the recurring decision of assigning workloads to 
nodes during cluster operation. In contrast, our work focuses on control-plane placement, a one-time infrastructure 
decision made during cluster initialisation. 

\subsubsection{Contextual Bandit Algorithms}
Contextual bandits are a class of reinforcement learning algorithms that learn to select the best action based on observed context, receiving immediate reward feedback after each decision. The LinUCB algorithm~\cite{li2010contextual} extends the classic Upper Confidence Bound (UCB) approach~\cite{auer2002} to contextual settings using linear function approximation. Zhou \textit{et al.}~\cite{zhou2020neural} introduced NeuralUCB, which integrates neural network function approximation with UCB-based exploration to achieve improved performance on complex, non-linear reward landscapes. While contextual bandits have been successfully applied to recommendation systems and online advertising, their application to infrastructure configuration problems remains largely unexplored. Our work bridges this gap by formulating control-plane placement as a contextual bandit problem, where infrastructure characteristics serve as context and operational performance metrics define rewards.


\begin{table}[t]
\centering
\caption{Summary of the closely related Work}
\label{tab:related_work}
\resizebox{\columnwidth}{!}{%
\begin{tabular}{|l|c|c|c|c|c|c|}
\hline
\textbf{Paper} & \textbf{Year} & \textbf{Focus} & \textbf{K3s} & \textbf{Control} & \textbf{Learning} & \textbf{Multi-} \\
 & & & & \textbf{Plane} & \textbf{Based} & \textbf{Region} \\
\hline
\hline
Koziolek \textit{et al.} \cite{koziolek2023} & 2023 & K8s Dist. Perf & \checkmark & \checkmark & -- & -- \\
\hline
Kjorveziroski \textit{et al.} \cite{kjorveziroski2022} & 2022 & Edge Serverless & \checkmark & -- & -- & -- \\
\hline
B{\"o}hm \& Wirtz \cite{bohm2021} & 2021 & Resource Usage & \checkmark & -- & -- & -- \\
\hline
{\v{C}}ili{\'c} \textit{et al.} \cite{cilic2023} & 2023 & Edge Containers & \checkmark & -- & -- & -- \\
\hline
Huang \textit{et al.} \cite{huang2020rlsk} & 2020 & K8s Scheduling & -- & -- & DRL & -- \\
\hline
Han \textit{et al.} \cite{han2021kais} & 2021 & K8s Edge-Cloud & -- & -- & MADRL & -- \\
\hline
Jian \textit{et al.} \cite{jian2024drs} & 2024 & Pod Scheduling & -- & -- & DRL & -- \\
\hline
Wang \textit{et al.} \cite{wang2023ppolrt} & 2023 & K3s Edge Sched. & \checkmark & -- & DRL & -- \\
\hline
Li \textit{et al.} \cite{li2010contextual} & 2010 & Bandits & -- & -- & CB & -- \\
\hline
Zhou \textit{et al.} \cite{zhou2020neural} & 2020 & Neural Bandits & -- & -- & CB & -- \\
\hline
\hline
\textbf{NL-CPS (Ours)} & \textbf{2025} & \textbf{CP Placement} & \checkmark & \checkmark & \checkmark & \checkmark \\
\hline
\end{tabular}%
}
\end{table}
\section{Methodology}\label{sec:problem}
\subsection{Problem Formulation}
As explained in Section~\ref{sec:swarmchestrate}, once Swarmchestrate identifies suitable resources for an application, these resources are assembled into a K3S cluster, where one node must be designated to host the control-plane. The control-plane comprises several critical components---including the API server, etcd datastore, scheduler, and controller manager---making its placement a performance-critical architectural decision. In this regard, the core problem addressed in this paper is that given a dynamically identified set of heterogeneous, geographically distributed nodes, which node should host the K3S control-plane to maximise cluster performance? 

We formalise this problem as a contextual bandit task, in which the system observes infrastructure characteristics (context), selects a host node (action), and receives operational performance feedback (reward) within a single decision cycle. Unlike sequential reinforcement learning---where actions influence future states and require multi-step credit assignment---control-plane placement is inherently a single-step decision. This means that once a node is selected, the cluster initialises, and the episode terminates without further state transitions. This episodic structure naturally aligns with the contextual bandit framework, which learns mappings from context to actions through repeated interactions without explicitly modelling temporal dependencies. 

More specifically, let the infrastructure comprise $N$ geographically distributed nodes $\mathcal{N} = \{n_1, n_2, \ldots, n_N\}$ deployed across multiple regions. Each node $n_i$ is characterised by a feature vector $\mathbf{f}_i = [c_i, m_i, \lambda_i]$ where $c_i$ denotes CPU cores, $m_i$ represents memory capacity in gigabytes (GBs), and $\lambda_i$ is the average network latency in milliseconds (ms) to peer nodes. The complete infrastructure context is represented as $\mathbf{x} = [\mathbf{f}_1, \mathbf{f}_2, \ldots, \mathbf{f}_N] \in \mathbb{R}^{N \times 3}$, yielding observation vectors of dimension 15, 24, 30, and 36 for the evaluated cluster sizes of 5, 8, 10, and 12 nodes respectively. All feature values are normalised using min-max scaling within each cluster instance to ensure balanced gradient contributions during neural network training. CPU capacity is discretised to realistic cloud instance configurations $c_i \in \{1, 2, 4\}$ cores, memory capacity is constrained to common allocation tiers $m_i \in \{1, 2, 4, 8\}$ GB, and network latency is sampled from the range $\lambda_i \in [10, 150]$ ms to capture both intra-regional and cross-region communication overhead.

The action space $\mathcal{A}$ is discrete and finite, comprising all possible node indices in the cluster such that $\mathcal{A} = \{a \mid a = 1, 2, \ldots, N\}$. Each action $a = i$ corresponds to the decision to deploy the K3S control-plane on node $n_i$, with the cardinality of the action space $|\mathcal{A}| = N$ varying proportionally with the size of the cluster. At each decision point, an agent observes context $\mathbf{x}$, selects action $a \in \{1, 2, \ldots, N\}$ corresponding to a candidate control-plane node, and receives reward $r(\mathbf{x}, a)$ quantifying the resulting cluster performance. The optimisation objective is:
\begin{equation}
    a^* = \arg\max_{a \in \{1,\ldots,N\}} \mathbb{E}[r(\mathbf{x}, a)]
    \label{eq:objective}
\end{equation}

The reward function $r: \mathcal{X} \times \mathcal{A} \rightarrow \mathbb{R}$ quantifies the quality of the placement of the control-plane based on the operational performance metrics measured. Upon selecting action $a_t = i$, the environment retrieves empirically measured K3S performance data corresponding to node $n_i$ operating as the control-plane under standardised benchmark workloads. The performance profile comprises six critical operational 
metrics, including, API server response latency $\ell_{\text{api}}$ in milliseconds, CPU utilisation of the control-plane node $u_{\text{cpu}}$ as a percentage, memory consumption $u_{\text{mem}}$ as a percentage, pod creation throughput $\phi_{\text{pod}}$ in pods per minute, pod creation average latency $\tau_{\text{pod}}$ in seconds, and pod deployment success rate $\rho_{\text{pod}}$ as the fraction achieving running state. The reward computation follows a penalty-bonus structure:
\begin{equation}
\begin{split}
r(\mathbf{x}, a) = R_{\text{base}} 
& - \text{Penalties}(\ell_{\text{api}}, u_{\text{cpu}}, u_{\text{mem}}, \tau_{\text{pod}}) \\
& + \text{Bonuses}(\phi_{\text{pod}}) - \text{Failures}(\rho_{\text{pod}})
\end{split}
\label{eq:reward}
\end{equation}
where the baseline reward $R_{\text{base}}$ is set to 100. Specifically, the reward computation applies continuous penalties for API latency ($-0.5$ per millisecond) and pod creation latency ($-5.0$ per second), reflecting that lower values are always preferable. Threshold-based penalties activate when resource utilisation exceeds operational limits, i.e. CPU utilisation above 85\% incurs $-3.0$ per percentage point over the threshold, and memory utilisation above 80\% incurs $-2.0$ per percentage point over the threshold. The bonus term rewards pod creation throughput 
at $+0.1$ per pod/min above a baseline of 100~pods/min, with no penalty for throughput below this baseline. Pod deployment failures incur penalties proportional to the failure rate, computed as 
$-100 \times (1 - \rho_{\text{pod}})$, ensuring the agent strongly prioritises cluster functionality over marginal performance optimisations. 
These thresholds align with Kubernetes operational best practices, where sustained CPU above 85\% or memory above 80\% typically indicates resource contention that degrades API server responsiveness.

\subsection{NL-CPS Framework}\label{sec:nlcps}
NL-CPS employs Neural LinUCB, a contextual bandit algorithm combining neural network function approximation with UCB exploration. Unlike sequential reinforcement learning methods such as Deep Q-Networks (DQN) or Proximal Policy Optimisation (PPO) that are designed for multi-step decision processes requiring experience replay buffers and target network synchronisation, contextual bandits perform direct supervised learning on observed context-action-reward tuples, making them computationally efficient and theoretically appropriate for single-step infrastructure decisions. 

\begin{figure}[t]
    \centering
    \includegraphics[width=0.49\textwidth]{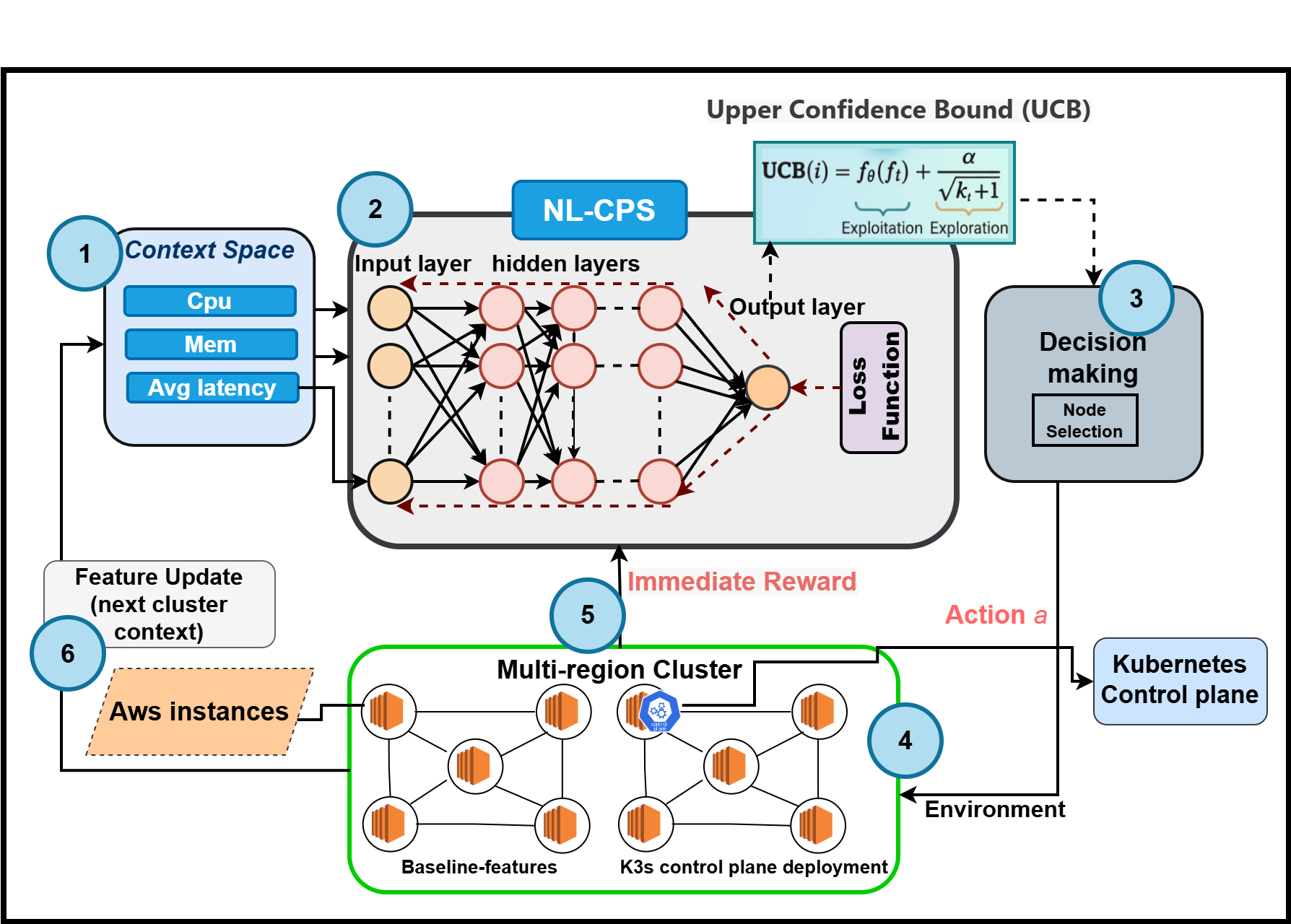}
    \caption{NL-CPS Architecture}
    \label{fig:nl_cps_architecture}
\end{figure}

Figure~\ref{fig:nl_cps_architecture} illustrates the overall NL-CPS architecture. In step~1, the context space captures infrastructure features, including CPU cores, memory capacity, and average network latency, for each candidate node in the cluster. These features are fed into the NL-CPS agent in step~2, which processes them through an input layer, hidden layers, and an output layer to produce expected reward estimates. In step~3, the UCB score is computed by combining the predicted reward with an exploration bonus, and the node with the highest score is selected for control-plane deployment. The selected action is applied to the environment in step~4, a multi-region K3S cluster comprising geographically distributed nodes. The environment evaluates the placement by measuring control-plane performance in step~5, and returns an immediate reward signal in step~6, which the agent uses to update its network parameters. This cycle repeats across successive training episodes, enabling the agent to progressively learn which node characteristics contribute to optimal control-plane performance.

NL-CPS maintains a neural network $f_\theta: \mathbb{R}^3 \rightarrow \mathbb{R}$ that predicts expected rewards for the features of the individual nodes. For each candidate node $i \in \{1, 2, \ldots, N\}$ with infrastructure features $\mathbf{f}_i$, the algorithm computes an UCB score:
\begin{equation}
    \text{UCB}(i) = \underbrace{f_\theta(\mathbf{f}_i)}_{\text{exploitation}} + \underbrace{\frac{\alpha}{\sqrt{k_i + 1}}}_{\text{exploration bonus}}
    \label{eq:ucb}
\end{equation}
where $f_\theta(\mathbf{f}_i)$ is the predicted reward of the neural network for node $i$ given its features, $\alpha = 0.5$ is a fixed hyperparameter that controls the intensity of the exploration, and $k_i$ is the selection count representing the number of times node $i$ has been selected during training. The UCB formulation mitigates the trade-off between exploration and exploitation, inherent in bandit problems. A purely exploitative agent that always selects the node with the highest predicted reward risks converging to a suboptimal choice if early predictions are inaccurate, whereas excessive exploration wastes training cycles on clearly inferior nodes. The exploration bonus $\alpha / \sqrt{k_i + 1}$ provides principled uncertainty quantification, being large for nodes with low selection counts (indicating high uncertainty in the reward estimate) and shrinking as nodes are repeatedly selected, thus indicating increased confidence. Since training proceeds over 10,000 timesteps across multiple sampled cluster configurations from the synthetic dataset, the selection count $k_i$ accumulates across episodes, tracking how frequently the agent has explored nodes with similar feature profiles. This mechanism ensures that under-explored regions of the feature space receive fair evaluation while well-characterised configurations compete primarily on predicted performance.

The reward prediction network comprises three fully-connected hidden layers containing 256, 256, and 128 neurons, respectively, each followed by Rectified Linear Unit (ReLU) activation functions. The output layer produces a single scalar representing the estimated expected reward for the input node features. This architecture depth enables the network to learn increasingly abstract representations of infrastructure characteristics. The early layers capture basic patterns such as the relationship between CPU capacity and memory size, while the deeper layers combine these to model complex interactions that influence control-plane performance. The network processes individual node feature vectors $\mathbf{f}_i \in \mathbb{R}^3$ rather than concatenated full contexts, allowing the model to generalise across different cluster sizes without retraining. Network parameters $\theta$ are optimised using the Adam optimiser with learning rate $3 \times 10^{-4}$.

Training proceeds through iterative interaction with the synthetic environment described in Section~\ref{sec:synthetic}. More importantly, training proceeds entirely within the synthetic environment and requires no physical cluster deployments during the learning phase; real K3S clusters are deployed only during the evaluation phase described in Section~\ref{sec:results}. 

Let $a_t \in \{1, \ldots, N\}$ denote the node selected at timestep $t$. At each timestep, the synthetic environment samples a cluster configuration from the pre-generated dataset, the agent computes UCB scores for all candidate nodes and selects the node with the highest score: $a_t = \arg\max_{i \in \{1,\ldots,N\}} \text{UCB}(i)$. The environment returns a reward $r_t$ computed from the synthetic performance model (not from actual pod creation), and the network parameters are updated via gradient descent: $\theta \leftarrow \theta - \eta \nabla_\theta (f_\theta(\mathbf{f}_{a_t}) - r_t)^2$. Finally, the selection count for the chosen node is incremented: $k_{a_t} \leftarrow k_{a_t} + 1$. The exploration parameter $\alpha$ remains fixed at 0.5 throughout training, providing a consistent exploration incentive. Since each training step involves only lightweight numerical operations (neural network forward pass, reward lookup, and backward pass) without physical infrastructure deployment, training proceeds for 10,000 timesteps per cluster configuration. Table~\ref{tab:hyperparameters} summarises the complete hyperparameter configuration used for NL-CPS training.

\begin{table}[t]
\centering
\caption{NL-CPS Hyperparameter Configuration}
\label{tab:hyperparameters}
\begin{tabular}{|l|c|}
\hline
\textbf{Parameter} & \textbf{Value} \\
\hline
\hline
Network architecture & [256, 256, 128] \\
\hline
Activation function & ReLU \\
\hline
Learning rate & $3 \times 10^{-4}$ \\
\hline
Exploration parameter ($\alpha$) & 0.5 \\
\hline
Training timesteps & 10,000 \\
\hline
Optimizer & Adam \\
\hline
\end{tabular}
\end{table}

\subsection{Synthetic Training Environment}
\label{sec:synthetic}
Training on physical infrastructure is impractical, as each K3S deployment cycle takes 15 to 20 minutes---meaning that 10,000 training iterations would require over 100 days. To overcome this, we developed a synthetic environment that models control-plane performance based on node characteristics, enabling rapid policy learning without the need for physical cluster deployments. This mainly includes the generation of two datasets, the baseline features and performance metrics. 

The baseline feature dataset was collected from a 5-node infrastructure deployed across geographically distributed regions, without initialising any K3S cluster. For each node, we recorded CPU core count, total memory capacity, and network latency to all peer nodes using ICMP round-trip measurements. This characterisation established the infrastructure feature space---comprising CPU capacity, RAM availability, and average network latency---that NL-CPS leverages as contextual information for placement decisions.

The performance metric dataset was obtained by deploying the K3S control-plane on each node individually. For each deployment, we measured API server latency, CPU and memory utilisation, pod creation throughput, and deployment success rate, then deleted the cluster and repeated the process with a different host node. These ground-truth measurements capture how control-plane performance varies with host node resource capacity and network positioning.

The measurements revealed consistent patterns such that throughput scales with CPU and memory capacity, nodes with limited resources exhibit 20 to 60\% deployment failure rates, and network latency reduces throughput proportionally. We encoded these patterns as deterministic functions that map node features $\mathbf{f}_i = [c_i, m_i, \lambda_i]$ to predicted performance metrics. Specifically, given a node's CPU cores, memory capacity, and average network latency, these functions output expected throughput, API latency, resource utilisation, and failure probability. Gaussian noise is added to the outputs to mitigate overfitting. Using this model, we generated 800 synthetic cluster configurations by sampling node features within observed ranges across 200 scenarios for each cluster size (5, 8, 10, and 12 nodes) to train our model. Importantly, the 12-node and 18-node clusters used for evaluation in Section~\ref{sec:results} are not part of this training data. The trained agent is evaluated on these unseen real-world deployments to assess whether policies learned from synthetic configurations generalise to new cluster sizes and topologies.

\subsection{Convergence Analysis} \label{sec:convergence}
\begin{figure}
    \centering
    \includegraphics[width=0.5\textwidth]{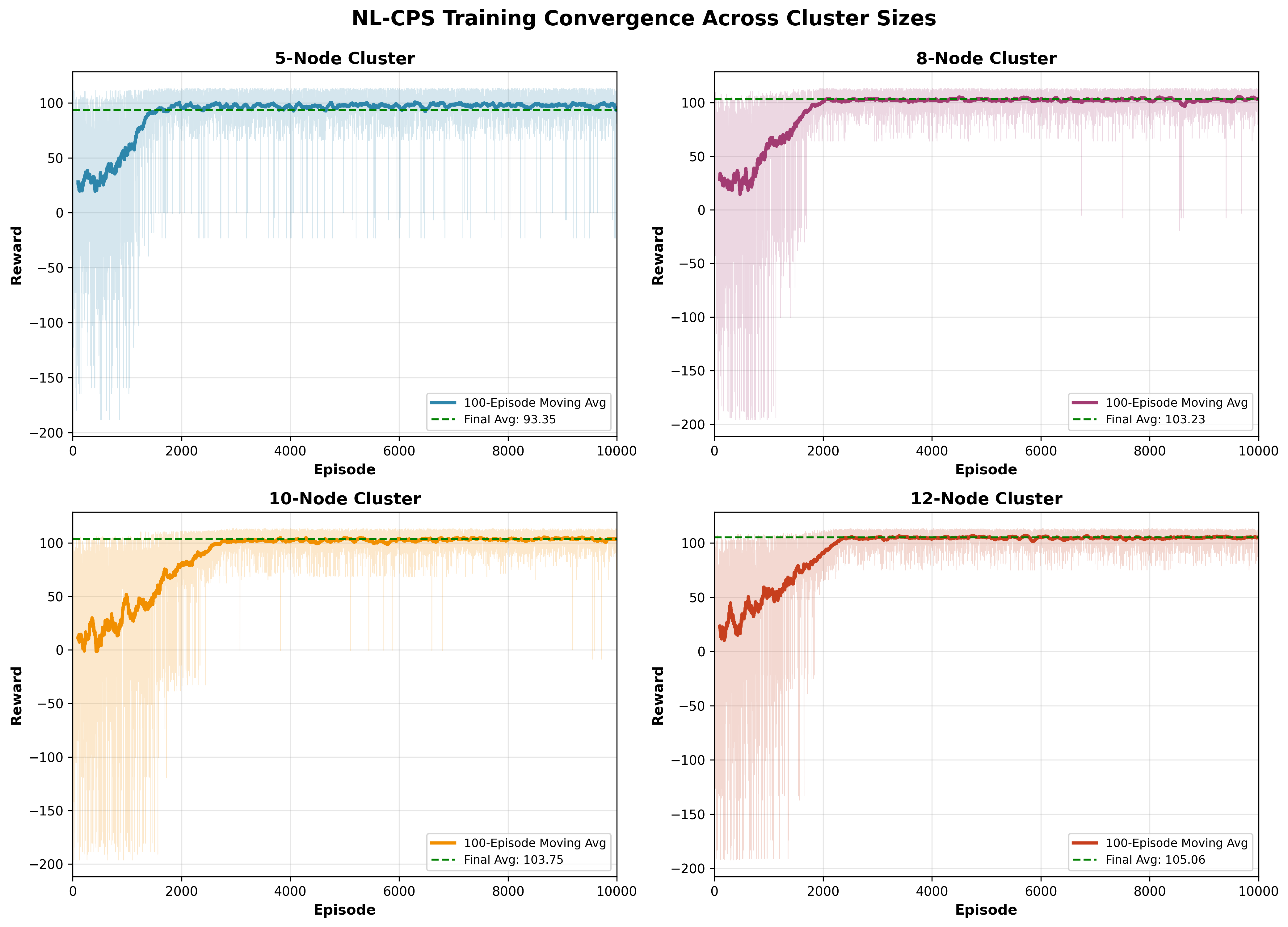}
    \caption{NL-CPS training convergence for 5, 8, 10, and 12-node 
synthetic clusters}
    \label{fig:nl_cps_convergence}
\end{figure}

Figure~\ref{fig:nl_cps_convergence} shows NL-CPS training performance across different cluster sizes. All configurations exhibit an initial exploration 
phase characterised by high reward variance, with instantaneous rewards occasionally dropping to $-200$ as the agent explores suboptimal node selections. The 5-node cluster demonstrates the fastest convergence, with the 100-episode moving average stabilising near the final value 
within approximately 1500 to 2000 timesteps. Larger clusters (8, 10, and 12 nodes) require additional exploration due to expanded action spaces, with convergence occurring around 2500 to 3000 timesteps. The shaded regions in Figure~\ref{fig:nl_cps_convergence} represent the reward variance, which decreases substantially as training progresses, indicating that the learned policy consistently identifies high-performing nodes with reduced uncertainty. Initial average rewards vary by cluster size, i.e. the 5-node configuration begins around 20 to 30, while larger clusters start closer to 0 due to the increased probability of 
selecting suboptimal nodes during early exploration. As training progresses, rewards rise steadily and stabilise between 95 and 105 across all configurations.

Final average rewards range from 93.35 for the 5-node configuration to 105.06 for the 12-node configuration, with larger clusters achieving slightly higher rewards due to increased node diversity enabling better placement optimisation. The 8-node (103.23) and 10-node (103.75) 
configurations achieve similar final performance, suggesting that beyond a threshold cluster size, additional nodes provide diminishing returns for placement optimisation.  The contextual bandit formulation is computationally efficient because each training step requires only one forward pass and one backward pass through the neural network. This avoids the additional overhead of experience replay buffers, target networks, and temporal difference updates that are typical of sequential reinforcement learning methods such as DQN and PPO.

\section{Experimental Setup and Results}\label{sec:evaluation}

\begin{table}[t]
\centering
\caption{Infrastructure nodes across geographic regions}
\label{tab:node_pool}
\scriptsize
\begin{tabular}{llcc}
\hline
\textbf{Region} & \textbf{Provider} & \textbf{Node IDs} & \textbf{Count} \\
\hline
London      & AWS    & 1, 2, 3    & 3 \\
Frankfurt   & AWS    & 17         & 1 \\
Ireland     & AWS    & 18         & 1 \\
Hungary     & SZTAKI & 5          & 1 \\
US-Virginia & AWS    & 4, 6, 7    & 3 \\
Hong Kong   & AWS    & 8, 9, 10   & 3 \\
Mumbai      & AWS    & 14         & 1 \\
Seoul       & AWS    & 15         & 1 \\
Tokyo       & AWS    & 16         & 1 \\
UAE         & AWS    & 11, 12, 13 & 3 \\
\hline
\multicolumn{3}{l}{\textbf{Total Nodes}} & \textbf{18} \\
\hline
\end{tabular}
\end{table}
\subsection{Infrastructure and Hardware Configuration} 
We provisioned a pool of 18 virtual machines across 10 geographic regions to enable flexible cluster composition. To capture realistic infrastructure heterogeneity, several deployment regions, including London, US-Virginia, Hong Kong, and the UAE, contain multiple nodes with varying specifications, while other regions, including Mumbai, Seoul, Tokyo, Frankfurt, Ireland, and Hungary, each contribute a single node. Table~\ref{tab:node_pool} provides the overall summary. Using this pool, we construct two evaluation clusters with $N \in \{12, 18\}$ nodes, with varying levels of geographic distribution and resource heterogeneity. All nodes run Ubuntu 22.04 LTS with memory capacity ranging from 1GB to 8GB and CPU capacity ranging from 1 to 4 vCPUs, reflecting typical heterogeneity in edge and cloud deployments where instance specifications differ 
by region and provider.


\begin{table}[t]
\centering
\caption{Infrastructure profile: 12-node multi-region cluster}
\label{tab:12node_infrastructure}
\scriptsize
\begin{tabular}{lcccc}
\hline
\textbf{Node} & \textbf{Region} & \textbf{CPU} & \textbf{RAM} & \textbf{Avg.\ Net} \\
 &  & \textbf{(cores)} & \textbf{(GB)} & \textbf{Lat.\ (ms)} \\
\hline
node1  & London      & 4 & 4 & 120.8 \\
node2  & London      & 2 & 1 & 110.1 \\
node3  & London      & 2 & 2 & 108.8 \\
node4  & US-Virginia & 1 & 2 & 129.9 \\
node5  & US-Virginia & 1 & 1 & 129.2 \\
node6  & US-Virginia & 4 & 8 & 129.5 \\
node7  & Hong Kong   & 1 & 1 & 144.7 \\
node8  & Hong Kong   & 2 & 2 & 143.5 \\
node9  & Hong Kong   & 2 & 1 & 143.0 \\
node10 & UAE         & 4 & 8 & 119.2 \\
node11 & UAE         & 2 & 2 & 118.5 \\
node12 & UAE         & 4 & 4 & 119.9 \\
\hline
\end{tabular}
\end{table}
\subsection{K-Bench Application and Workload Configuration}
To evaluate how control-plane placement impacts cluster performance, we employed k-bench \cite{k-bench}, an open-source Kubernetes benchmarking tool. K-bench is a configurable framework for stress-testing Kubernetes clusters, issuing CRUD operations against the API server via its integrated Kubernetes client, and enabling systematic evaluation of API server responsiveness and pod scheduling efficiency. We configure k-bench with 12 concurrent clients to emulate a moderately loaded control-plane representative of typical edge deployment scenarios, allowing direct comparison between the NL-CPS and baseline placement strategies under identical workload conditions.

Three workload configurations are evaluated to assess performance across varying intensity levels: 1) a light workload comprising 24 individual pod creations, 2) a medium workload deploying 40 pods to assess moderate concurrency, and 3) a heavy workload deploying 120 pods in groups of five replicas to stress-test sustained throughput. Additionally, we measure CRUD operation latencies for namespace and service objects to isolate API server responsiveness from scheduler overhead. This multi-workload evaluation enables us to determine whether NL-CPS maintains performance advantages as operational load increases. Each experiment is repeated 10 times, and the average result with standard deviation is reported.
\subsubsection{Baseline Selection Strategies}
To evaluate NL-CPS, we compare it against three baseline strategies, each reflecting a distinct placement philosophy. \textbf{HIGH-RES} selects the node with the maximum computational resources but higher network latency, testing whether abundant CPU and memory capacity can offset suboptimal network positioning. \textbf{LOW-LATENCY} selects the node with the lowest average network latency to peer nodes, evaluating the hypothesis that control-plane performance depends primarily on communication efficiency with worker nodes. Lastly, \textbf{RANDOM} selects a node at random, emulating the implicit behaviour of kubeadm when control-plane placement is not explicitly configured. The inclusion of both resource-centric (HIGH-RES) and network-centric (LOW-LATENCY) baselines enables direct assessment of whether NL-CPS learns meaningful trade-offs between competing optimisation objectives, rather than simply maximising a single infrastructure dimension.

\subsection{Results and Discussion}\label{sec:results}
\subsubsection{12-Node Cluster}
The 12-node cluster (Table~\ref{tab:12node_infrastructure})  represents a geo-distributed deployment spanning four AWS regions: Europe (London), North America (US-Virginia), Asia (Hong Kong),
and the UAE. This configuration introduces greater infrastructure heterogeneity with CPU capacity ranging from 1 to 4 vCPUs, 1 to 8GB memory, and average network latencies of 108 to 143~ms. NL-CPS, based on the learned policy, selects \textbf{node10 (UAE)} as the control-plane host, combining high computational resources (4~vCPU, 8~GB memory) with moderate network latency (119.2~ms). Notably, NL-CPS did not select node3 (London), which offers the lowest network latency (108.8~ms) but limited resources (2~vCPU, 2~GB), nor node6 (US-Virginia), which matches node10's hardware specification but exhibits 10.3~ms higher latency (129.5~ms). This selection demonstrates that the learned policy balances both resource capacity and network positioning rather than optimising either dimension in isolation.

\begin{figure}[t]
    \centering
    \includegraphics[width=0.48\textwidth]{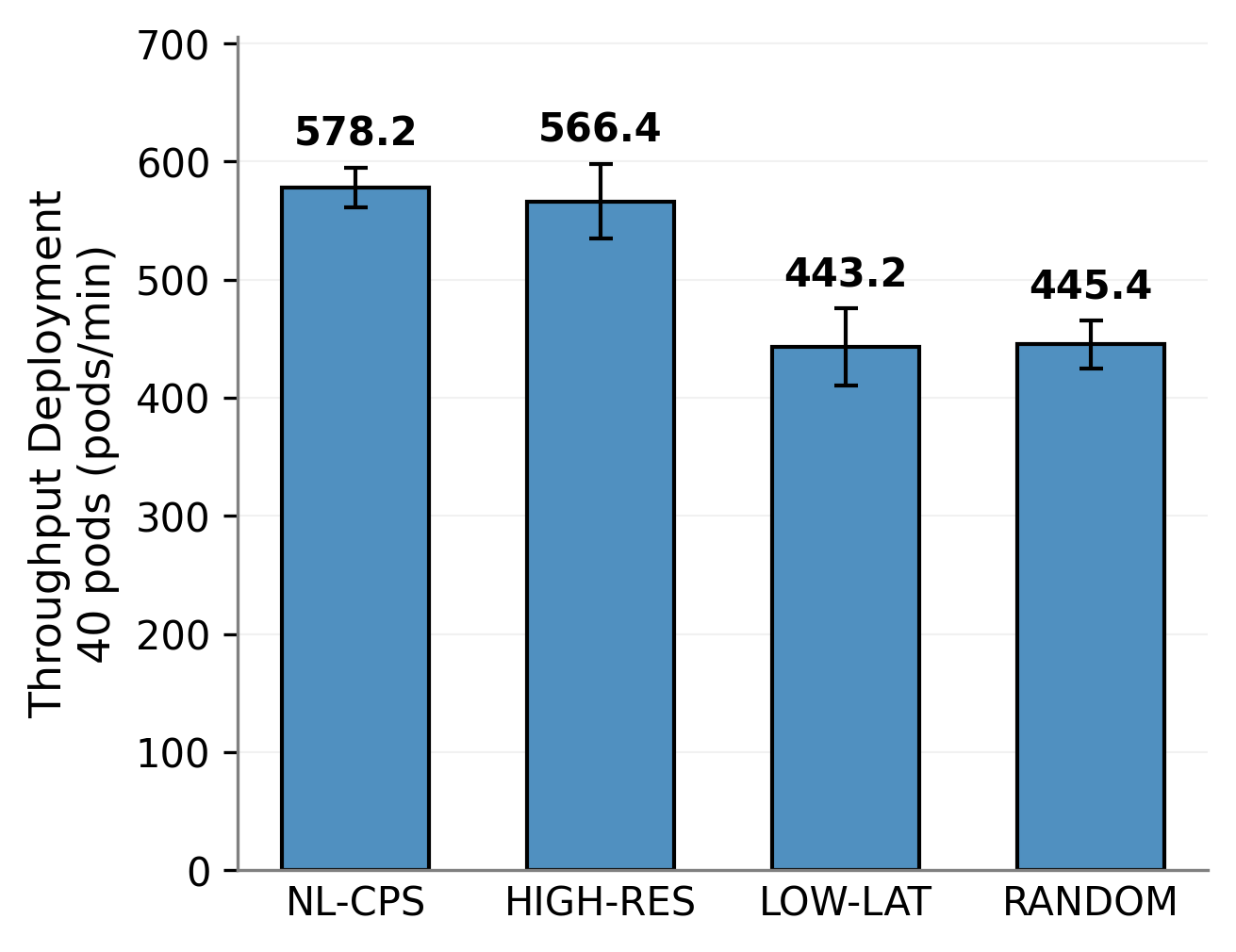}
    \caption{12-node cluster: 40-pod deployment throughput}
    \label{fig:12node_throughput}
\end{figure}

\begin{figure}[t]
    \centering
    \includegraphics[width=0.4\textwidth]{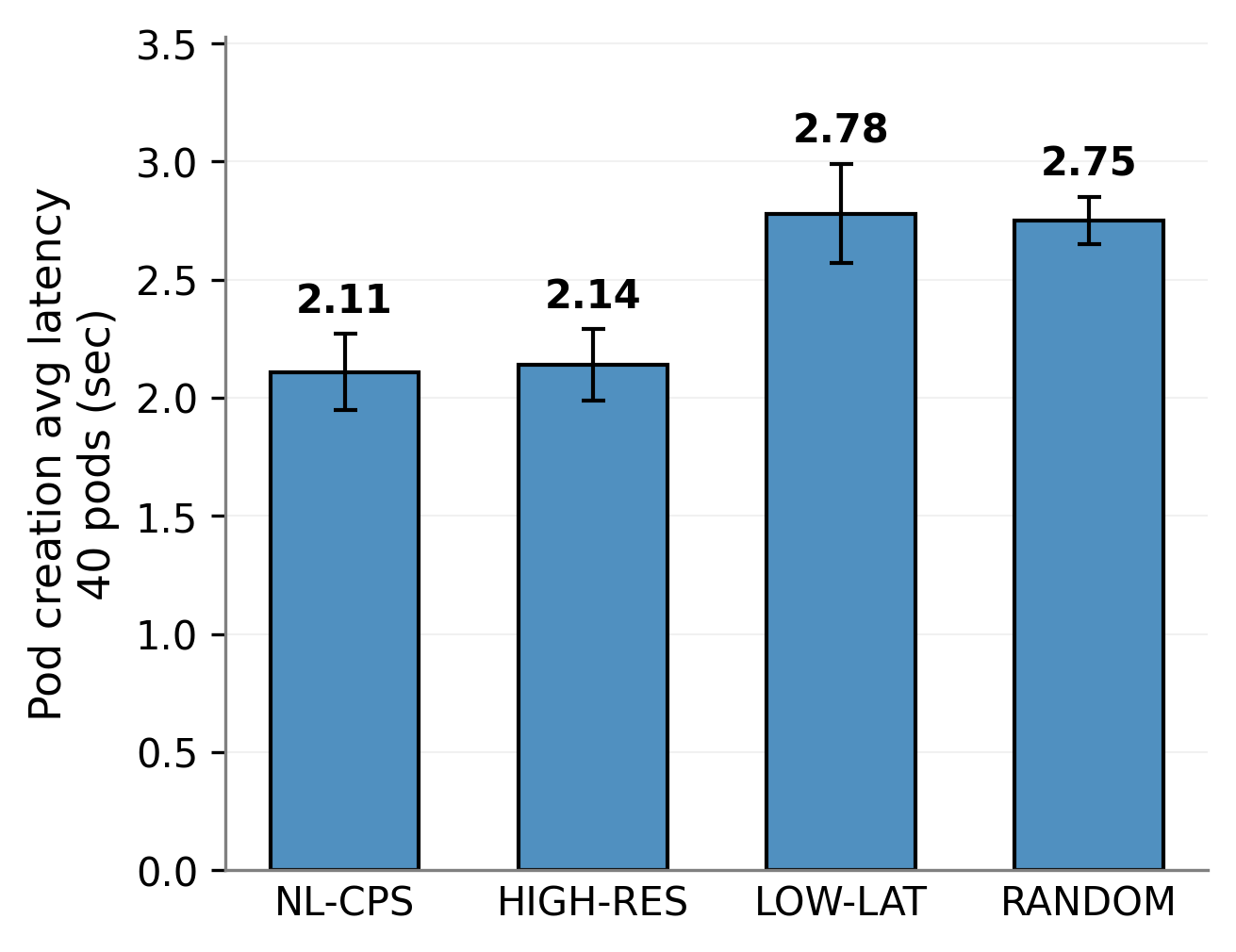}
    \caption{12-node cluster: Pod creation latency (40-pod deployment). 
    NL-CPS achieves 24.1\% reduction compared to LOW-LATENCY.}
    \label{fig:12node_pod_latency}
\end{figure}

\begin{table}[t]
\centering
\caption{Infrastructure profile: 18-node multi-region cluster}
\label{tab:18node_infrastructure}
\scriptsize
\begin{tabular}{lccccc}
\hline
\textbf{Node} & \textbf{Region} & \textbf{Provider} & \textbf{CPU} & \textbf{RAM} & \textbf{Avg.\ Net} \\
 &  &  & \textbf{(cores)} & \textbf{(GB)} & \textbf{Lat.\ (ms)} \\
\hline
node1  & London      & AWS    & 4 & 8 & 120.8 \\
node2  & London      & AWS    & 1 & 1 & 110.1 \\
node3  & London      & AWS    & 2 & 2 & 108.8 \\
node4  & US-Virginia & AWS    & 1 & 2 & 129.9 \\
node5  & Hungary     & SZTAKI & 4 & 4 & 108.6 \\
node6  & US-Virginia & AWS    & 2 & 2 & 129.2 \\
node7  & US-Virginia & AWS    & 4 & 8 & 129.5 \\
node8  & Hong Kong   & AWS    & 2 & 1 & 144.7 \\
node9  & Hong Kong   & AWS    & 2 & 2 & 143.5 \\
node10 & Hong Kong   & AWS    & 2 & 4 & 143.0 \\
node11 & UAE         & AWS    & 4 & 8 & 119.2 \\
node12 & UAE         & AWS    & 2 & 2 & 118.5 \\
node13 & UAE         & AWS    & 2 & 4 & 119.9 \\
node14 & Mumbai      & AWS    & 2 & 2 & 115.1 \\
node15 & Seoul       & AWS    & 2 & 2 & 164.9 \\
node16 & Tokyo       & AWS    & 2 & 2 & 157.1 \\
node17 & Frankfurt   & AWS    & 4 & 8 & 110.2 \\
node18 & Ireland     & AWS    & 2 & 2 & 120.2 \\
\hline
\end{tabular}
\end{table}

Figure~\ref{fig:12node_throughput} presents deployment throughput for the 40-pod workload. NL-CPS achieves the highest mean throughput of 578.2~pods/min, outperforming the LOW-LATENCY variant by 30.5\% (443.2~pods/min) and RANDOM by 29.8\% (445.4~pods/min). Since NL-CPS and HIGH-RES run on identical hardware (4~vCPU, 8~GB), the 2.1\% throughput difference is attributable solely to NL-CPS's 10.3~ms lower average network latency to the worker nodes. Figure~\ref{fig:12node_pod_latency} presents pod creation latency, measuring the time from API request to pod reaching running state. NL-CPS achieves 2.11~seconds per pod, representing a 24.1\% reduction compared to the 2.78~s of LOW-LATENCY and 23.3\% compared to 2.75~s of RANDOM. Despite possessing minimum network latency (108.8~ms), LOW-LATENCY exhibits the highest per-pod latency, confirming that resource constraints on node3 (2~vCPU, 2~GB) create processing
bottlenecks that network proximity cannot mitigate. The similarity between NL-CPS (2.11~s) and HIGH-RES (2.14~s) reflects their identical computational capacity, with the modest 1.4\% difference attributable to network positioning.

\begin{figure}[t]
    \centering
    \includegraphics[width=0.4\textwidth]{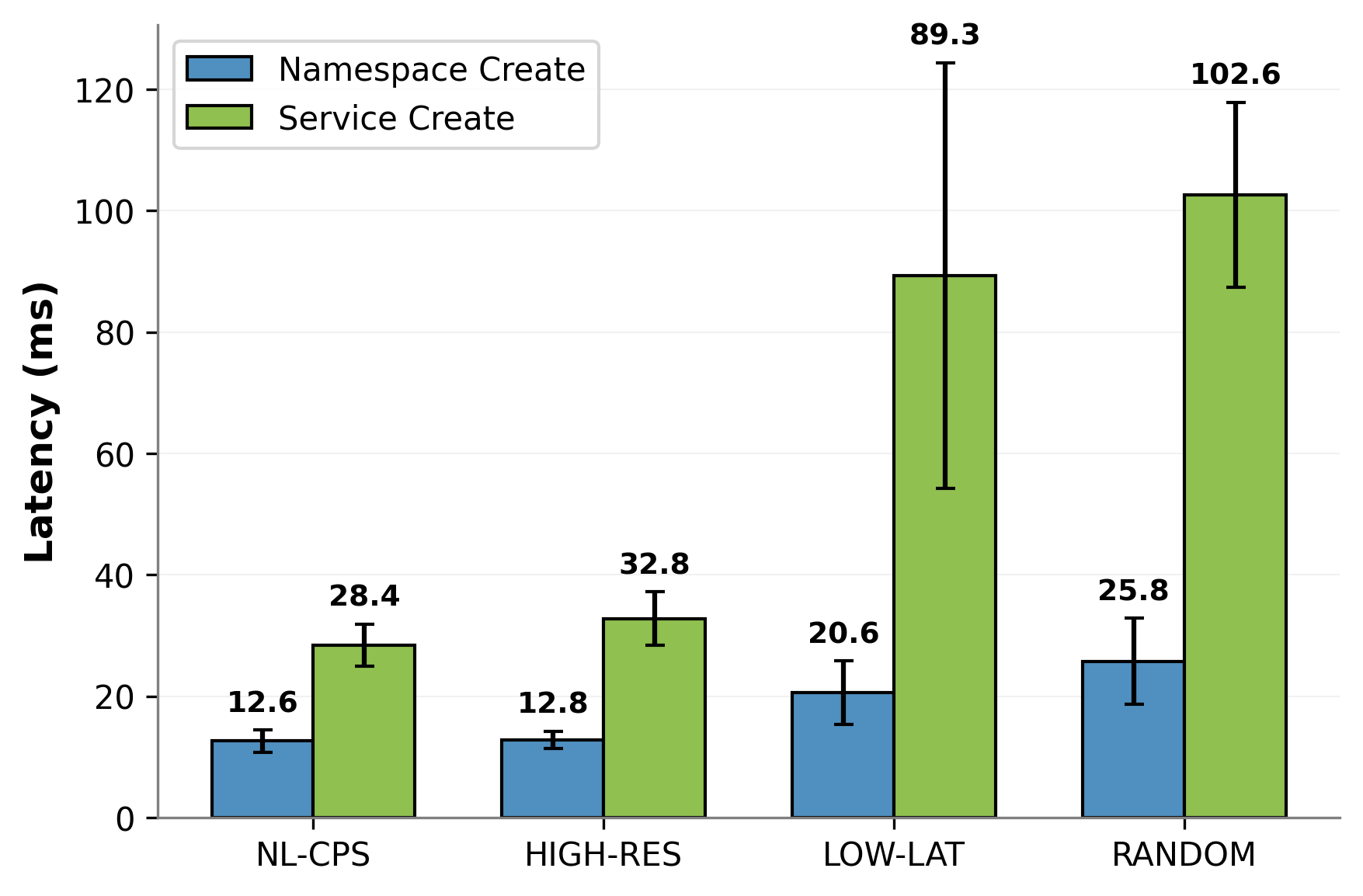}
    \caption{12-node cluster: CRUD operation latencies. NL-CPS achieves 
    72.3\% reduction in service creation compared to RANDOM.}
    \label{fig:12node_crud}
\end{figure}

To isolate API server responsiveness from scheduler overhead, we measured CRUD operation latencies (Figure~\ref{fig:12node_crud}). NL-CPS achieves namespace creation latency of 12.6~ms, representing a 51.2\% reduction compared to 25.8~ms of RANDOM. Service creation shows even greater differentiation with an overall 72.3\% reduction, with NL-CPS achieving 28.4 ms versus 102.6 ms for RANDOM. LOW-LATENCY exhibits elevated write latencies (89.3~ms for service creation) despite competitive read performance, indicating that write-intensive operations are particularly sensitive to computational capacity. The 12-node results validate the NL-CPS scalability to larger heterogeneous deployments. Across all metrics, NL-CPS consistently outperforms single-objective heuristics, demonstrating that the learned placement policy effectively balances resource capacity and network proximity, even in previously unseen cluster topologies.
\subsubsection{18-Node Cluster}
The 18-node cluster (Table~\ref{tab:18node_infrastructure}) represents our largest experimental deployment, spanning ten geographic regions across Europe (London, Frankfurt, Ireland, Hungary), North America (US-Virginia), the UAE, and Asia-Pacific (Hong Kong, Mumbai, Seoul, Tokyo). This configuration introduces infrastructure heterogeneity, with CPU capacity ranging from 1 to 4 vCPUs, 1 to 8GB memory, and average network latencies from 108 to 165ms. The cluster contains four nodes with maximum resources (4 vCPUs, 8GB memory): node1 (London, 
120.8~ms), node7 (US-Virginia, 129.5~ms), node11 (UAE, 119.2~ms), and node17 (Frankfurt, 110.2~ms), highlighting the scenario where hardware-equivalent candidates differ only in network positioning.

NL-CPS, based on the learned policy, selects node17 (Frankfurt) as the control-plane host, combining high computational resources (4~vCPU, 8GB memory) with the lowest network latency among high-resource nodes (110.2~ms). HIGH-RES selects node1 (London, 120.8 ms latency, 4 vCPU, 8 GB RAM), matching the hardware capacity chosen by NL-CPS but with 10.6 ms higher network latency. LOW-LATENCY selects node5 (Hungary, 108.6 ms latency, 4 vCPU, 4 GB RAM), providing the lowest network latency; however, with reduced memory capacity. RANDOM selects node15 (Seoul, 164.9 ms latency, 2 vCPU, 2 GB RAM), a placement with both constrained resources and elevated latency.

\begin{figure}[t]
    \centering
    \includegraphics[width=0.45\textwidth]{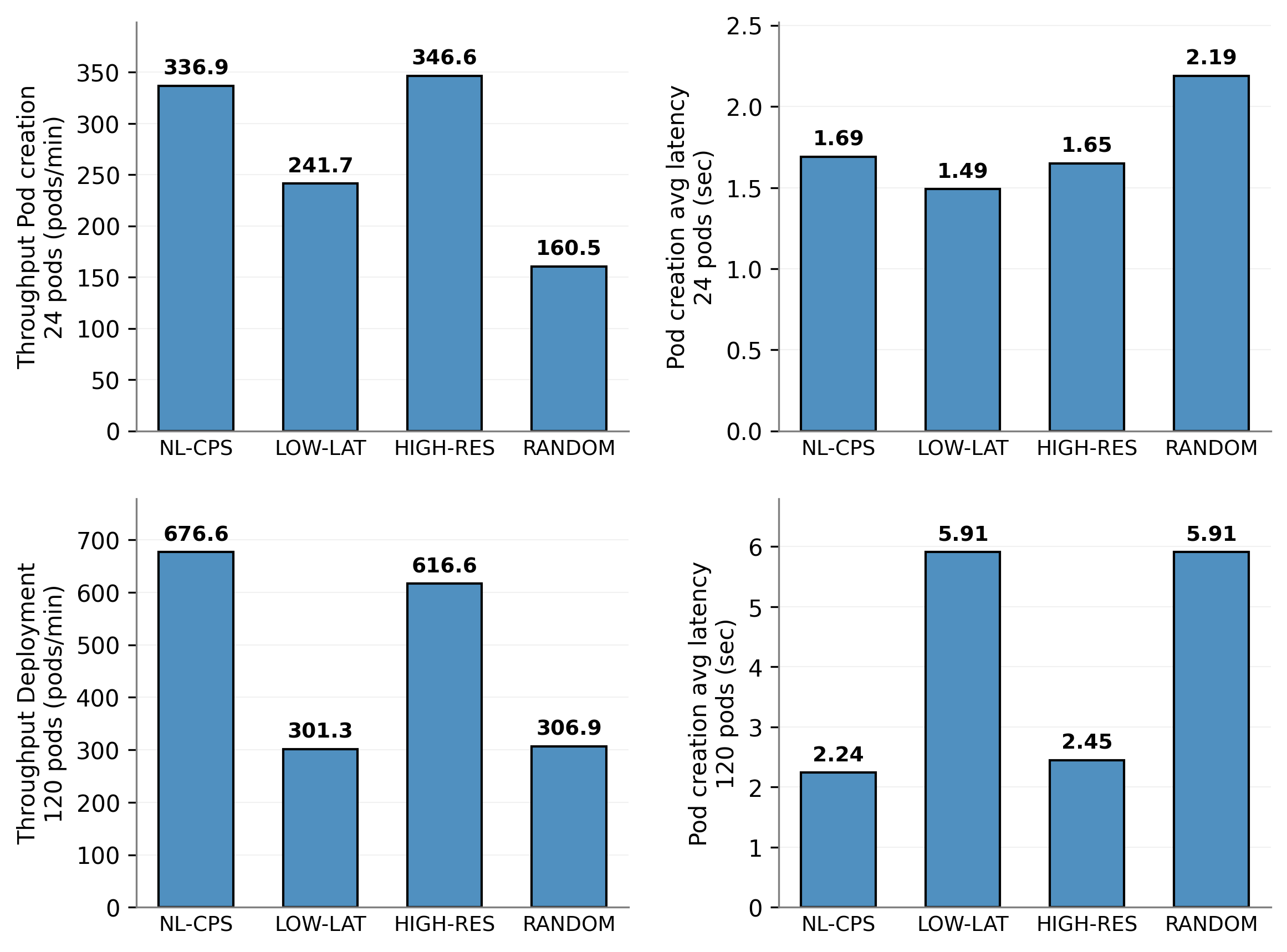}
    \caption{18-node cluster: Throughput and latency across placement 
    strategies for 24-pod and 120-pod workloads.}
    \label{fig:18node_results}
\end{figure}

Figure~\ref{fig:18node_results} reports the results across the different workload intensities. For the 24-pod workload, HIGH-RES achieves marginally higher throughput (346.6~pods/min) than NL-CPS (336.9~pods/min), while LOW-LATENCY achieves lower per-pod latency (1.49~s vs 1.69~s). These results reflect that under light load, HIGH-RES and LOW-LATENCY each provide localised advantages when their respective bottlenecks are not stressed.

In contrast, under heavy workload, e.g. in the 120-pod case, NL-CPS demonstrates superiority across all metrics. NL-CPS achieves 676.6~pods/min throughput, outperforming HIGH-RES by 9.7\% (616.6~pods/min) and LOW-LATENCY by 124.6\% (301.3~pods/min). NL-CPS and HIGH-RES share identical hardware (4 vCPU, 8 GB RAM), so the 9.7\% throughput difference is attributable entirely due to NL-CPS's 10.6~ms lower network latency, which accumulates over hundreds of API round-trips during sustained pod deployment.

The LOW-LATENCY baseline exhibits severe performance degradation under heavy workload despite possessing the minimum network latency (108.6~ms) and larger CPU capacity (4~vCPU). The LOW-LATENCY node has 4GB RAM compared to 8GB in the case of NL-CPS. Under a light workload, this difference has minimal impact; however, under heavy load, the reduced memory headroom causes the API server to queue requests as control-plane processes compete for resources during concurrent pod scheduling. Pod creation latency increases from 1.49 s (24 pods) to 5.91 s (120 pods) for the baseline, representing a 297\% degradation, whereas NL-CPS exhibits a more gradual behaviour with an increase from 1.69 s to 2.24 s (33\%). This asymmetric degradation confirms that memory capacity becomes the dominant bottleneck under sustained control-plane load.

The 18-node results validate NL-CPS scalability to production-scale deployments spanning multiple geographic regions. The selection of node17 over three other hardware-equivalent candidates (node1, node7, node11) demonstrates that NL-CPS correctly discriminates based on network latency when computational resources are held constant. Across heavy workloads, NL-CPS consistently outperforms single-objective heuristics, confirming that learned placement policies effectively balance resource adequacy with network proximity in complex, previously unseen cluster topologies.
\section{Conclusion} \label{sec:Conc}
This paper presented NL-CPS, a Neural LinUCB-based contextual bandit framework for optimal Kubernetes control-plane placement in heterogeneous multi-region K3S clusters. By formalising control-plane placement as a single-step decision problem, we established contextual bandits as the theoretical framework, distinguishing our approach from sequential reinforcement learning methods designed for recurring scheduling decisions. NL-CPS learns placement policies directly from infrastructure context, including CPU capacity, memory availability, and network latency, while balancing exploitation and exploration through UCB scoring. Experimental evaluation demonstrates that NL-CPS consistently outperforms single-objective baselines, confirming that control-plane performance is governed by a joint interaction of memory capacity and network positioning that neither resource-centric nor latency-centric heuristics capture in isolation. Future work will extend NL-CPS to support dynamic control-plane migration during runtime and high-availability configurations.
\section*{Acknowledgments}
This work is co-funded by the European Union’s Horizon Europe programme under grant agreement No. 101135012, and by UK Research and Innovation (UKRI) under grant agreement No. 10102651, as part of the project Swarmchestrate: Application-level Swarm-based Orchestration Across the Cloud-to-Edge Continuum.
\bibliographystyle{IEEEtran}
\bibliography{Ref}
\end{document}